\documentclass[prl,aps,twocolumn,showpacs,preprintnumbers,amsmath,amssymb,
superscriptaddress,notitlepage]{revtex4}

\usepackage{graphicx}
\usepackage{dcolumn}
\usepackage{bm}
\usepackage{hyperref}

scaled\magstep1

\let\a=\alpha \let\b=\beta    \let\d=\delta \let\e=\varepsilon
  \let\h=\eta   \let\th=\theta \let\k=\kappa \let\l=\lambda
\let\m=\mu    \let\n=\nu             \let\r=\rho
\let\s=\sigma     
   \let\o=\omega
\let\G=\Gamma \let\D=\Delta  \let\L=\Lambda 
         
\let\O=\Omega 

 \def\VV{{\cal V}}
 \def\WW{{\cal W}}
 \def\BBB{{\cal B}}
\def\RR{{\cal R}}\def\LL{{\cal L}}

   \def\pp{{\bf p}}
 \def\xx{{\bf x}} \def\yy{{\bf y}} 

\def\kk{{\bf k}}

\def\\{\hfill\break}
\let\==\equiv
\let\io=\infty
\let\0=\noindent

\def\tende#1{\,\vtop{\ialign{##\crcr\rightarrowfill\crcr\noalign{\kern-1pt
    \nointerlineskip} \hskip3.pt${\scriptstyle #1}$\hskip3.pt\crcr}}\,}
\def\otto{\,{\kern-1.truept\leftarrow\kern-5.truept\to\kern-1.truept}\,}

\def\to{\rightarrow}
\def\la{\left\langle}
\def\ra{\right\rangle}
\def\qed{\hfill\raise1pt\hbox{\vrule height5pt width5pt depth0pt}}

\def\Tr{{\rm Tr}}

\def\be{\begin{equation}}
\def\ee{\end{equation}}
\def\bea{\begin{eqnarray}}
\def\eea{\end{eqnarray}}
\def\nn{\nonumber}
\def\pref#1{(\ref{#1})}

\begin{document}

\title{Conductivity between Luttinger liquids:
coupled chains and bilayer graphene}
\author{Vieri Mastropietro}%
\affiliation{%
Universit\`a di Roma Tor Vergata, Viale della Ricerca Scientifica
00133 Roma, Italy,EU}

\begin{abstract}
The conductivity properties between Luttinger liquids are analyzed
by exact Renormalization Group methods. We prove that in a two
chain system or in a model of bilayer graphene, described by two
coupled fermionic honeycomb lattices interacting with a gauge
field, the transverse optical conductivity at finite temperature
is anomalous and decreasing together with the frequency as a power
law with Luttinger liquid exponent.
\end{abstract}
 \maketitle

\section{Introduction}

The many body interaction in fermionic systems can destroy the
electron-like nature of the elementary excitations, a fact which
can have deep consequences on the transport properties. This was
pointed out first by Anderson \cite{A1}, who got evidence that the
coherent transport between Luttinger liquids due to particle
hopping is strongly depressed with respect to the non interacting
case.
One of the first applications of this idea was, see \cite{A2,A2a},
an explanation of the c-axis anomalous conductivity between planes
in high $T_c$ superconductors, with the {\it assumption} of
Luttinger liquid behavior in the planes described by 2D square
lattice Hubbard models; while intriguing, this theory suffers the
fact that no convincing theoretical evidence has been found up to
now to substantiate such assumption. Subsequently, the attention
was focused to one dimensional systems, where Luttinger liquid
behavior is surely present; in addition to a theoretical interest,
the main physical motivation was that coupled fermionic chains
well describe quasi-one dimensional organic conductors, see {\it
e.g.} \cite{8}. At zero temperature Renormalization Group or
bosonization analysis apparently indicate that the hopping can
destroy Luttinger liquid behavior in several regimes, see {\it
e.g.} \cite{0,0a,10,1,1a,1b,2,2a,2b,2c,3,3a,M}. On the other hand
at higher temperatures the system still shows Luttinger liquid
properties \cite{9,9a,9c,5}.
The transverse conductivity between weakly coupled fermionic
chains at temperatures and frequencies greater than the hopping
was shown to be $\s^{\perp}(\o)\sim \o^{\a}$ with $\a$ related to
the Luttinger liquid exponent $\h$; in particular the value of the
exponent $\a$ was claimed to be $\a=2\h$, in \cite{4}, where a
tunnelling approach was followed, and in \cite{6}, by using the
Kubo formula; on the other hand, in \cite{7} the {\it different}
value $\a=2\h-1$ was found, by using dynamical mean field theory.
The reason of these discrepancies relies on the fact that the
computations were done in the so called Tomonaga-Luttinger or {\it
g-ology} approximation, in which the fermions close to the two
Fermi points are described in terms of massless Dirac particles.
This simplifies the computations and allows the use of powerful
techniques like bosonization, but introduces spurious ultraviolet
divergences in the conductivity which therefore needs a
regularization, and different regularizations produces different
results (see {\it e.g.} the discussion in \cite{7} after eq.(3)).

In conclusion, there are at present no firm results on the
conductivity between higher dimensional Luttinger liquids, and
even in one dimension there are still ambiguities due to
approximations and regularizations. In this paper the conductivity
properties between Luttinger liquids at finite temperature are
analyzed by the {\it exact} Renormalization Group methods
developed starting from \cite{BG}; indeed the suggestion of using
such techniques for this problem dates back to Anderson himself
\cite{A1} but their full development required a long time. In the
case of coupled spinless chains we get an exact expression for the
transverse conductivity (the lattice furnishes the natural
cut-off), which is, at temperatures and frequencies greater than
the hopping
%
\be \s^{\perp}_\b(\o_n)\sim t^2 \o_n^{2\h-1}\label{x1}\ee
if $\o_n={2\pi\over\b}(n+{1\over 2})$. The analysis is based on
the implementation of Ward Identities in the Renormalization
Group, with rigorous bounds for the corrections due to the lattice
\cite{BM}. Note also that the computation of the parallel
conductivity gives in this regime $\s^{\parallel}_\b(\o_n)\sim
\o_n^{-1}$, that is no anomalous exponent appears in the frequency
dependence in that case.

In addition to justifying the anomalous exponent predicted by mean
field theory in \cite{7} for the two chain problem, the exact
Renormalization Group methods can provide for the first time
evidence for anomalous transverse conductivity between {\it
bidimensional} Luttinger liquids, for which bosonization cannot be
applied. Electrons on the honeycomb lattice interacting with an
$U(1)$ gauge field, representing the retarded e.m. interaction or
the effect of disorder or ripples \cite{V2}, have Luttinger liquid
behavior. Indeed this system was first analyzed in the continuum
Dirac approximation in the early work \cite{GGV}, where evidence
of Luttinger liquid was found based on second order perturbation
theory. Later on, in \cite{GMP1,GMP2} Luttinger liquid behavior
was established {\it at any order} and taking rigorously into
account the honeycomb lattice, by implementing lattice Ward
Identities in the Renormalization Group scheme. The behavior of
the two point function is similar to the one of the spinless
chain; the wave function renormalization has a power law with
exponent $\h$. By coupling two interacting fermionic honeycomb
lattices by an hopping term we get a model for {\it bilayer
graphene} \cite{V3}. The zero temperature properties of such
system are rather complex and still not completely understood, see
{\it e.g.} \cite{V4,V5,V6} and the review \cite{V7}. However, as
in the case of coupled chains, at finite temperature and
frequencies the Luttinger liquid behavior of the uncoupled system
can reveal itself by the transverse conductivity; in particular we
will show that for temperatures and frequencies greater than the
hopping
%
\be\s^\perp_\b(\o_n)\sim t^2 \o_n^{2\h}\label{x2}\ee
while is essentially constant in the non interacting case. This
confirms for the first time in two dimensions the Anderson idea:
coherent transport between Luttinger liquids is depressed with
respect to non interacting systems or Fermi liquids. Moreover, the
presence of an anomalous Luttinger liquid exponents in the
frequency dependence of the transverse conductivity of bilayer
graphene could be revealed in future experiments.

The paper is organized in the following way. In \S II we derive
the transverse conductivity for the two chain model; in \S III we
derive the transverse conductivity for bilayer graphene. In App A
the computation of the conductivity in the non interacting case is
presented.

\section{The two chain model}

We consider a two chain model described as two one dimensional
interacting spinless fermionic systems coupled by an hopping term;
the Hamiltonian is
\be H=H_1+H_2+P\label{1} \ee
where, for $i=1,2$, $H_i=H_i^{(0)}+V_i$
\bea &&H_i^{(0)}=-{1\over 2}\sum_{x=1}^{L-1}
(a^+_{x+1,i}a^-_{x,i}+a^+_{x,i}a^-_{x+1,i})\nn\\
&&V_i=-\l\sum_{x,y=1}^{L-1} v(x-y)a^+_{x,i}a^-_{x,i}
a^+_{y,i}a^-_{y,i}\eea
and
\be P=-t\sum_{x=1}^{L-1}
[a^+_{x,1}a^-_{x,2}+a^+_{x,2}a^-_{x,1}]\ee
where $a^\pm_{x,i}$ are fermionic operators and $|v(x)|\le e^{-\k
|x|}$. Either $\l$ and $t$ are assumed to be {\it small}.

%
%
%
%
%
%
%
As usual we can introduce the interaction with an e.m. field with
a Peierls substitution $H\to H(A)$ with $V_i(A)=V$ and, if
$A=(A^{\parallel},A^{\perp})$,
\be H_i^{(0)}(A)=-\sum_x {1\over 2}
(a^+_{x+1,i}e^{iA_{x}^{\parallel}}a^-_{x,i}+a^+_{x,i}e^{-iA_{x}^{\parallel}}a^-_{x+1,i})
\ee
and
\be P(A)=-t\sum_x
(a^+_{x,1}e^{iA^{\perp}_{x}}a^-_{x,2}+a^+_{x,2}e^{-iA^{\perp}_{x}}a^-_{x,1})\ee
The {\it parallel current} is defined as
\be {\partial H(A)\over\partial
A^{\parallel}_x}=j^{P,\parallel}_{x}+A_x^{\parallel}
j^{D,\parallel}_{x}+O((A^{\parallel})^2)\ee
where $j^{P,\parallel}_{x,i}$ and $j^{D,\parallel}_{x,i}$ are
called respectively {\it paramagnetic} and {\it diamagnetic} part
of the current and are given by
\bea j^{P,\parallel}_{x}=\sum_{i=1}^2{1\over
2 i}(a^+_{x+1,i}a^-_{x,i}-a^+_{x,i}a^-_{x+1,i})\nn\\
j^{D,\parallel}_{x}=\sum_{i=1}^2 {1\over
2}(a^+_{x+1,i}a^-_{x,i}+a^+_{x,i}a^-_{x+1,i})\eea
The {\it transverse current} is defined as
\be {\partial H(A)\over\partial
A^{\perp}_x}=j^{P,\perp}_{x}+A_x^{\perp}j^{D,\perp}_{x,i}+O((A^{\perp})^2)\ee
where $j^{P,\perp}_{x}$ and $j^{D,\perp}_{x}$ are called
respectively {\it paramagnetic} and {\it diamagnetic} part of the
current and are given by \bea
&&j^{P,\perp}_{x}={t\over i}(a^+_{x,1}a^-_{x,2}-a^+_{x,2}a^-_{x,1})\nn\\
&&j^{D,\perp}_{x}=t(a^+_{x,1}a^-_{x,2}+a^+_{x,2}a^-_{x,1}) \eea
Finally the fermionic density is $\r_x= a^+_{1,x}
a_{1,x}+a^+_{2,x} a_{2,x}$.

%
%

If $\pp=(\o_n,p)$, $\o_n={2\pi\over\b}n, p={2\pi\over L}m$, the
transverse conductivity at finite temperature is given by
\be \s^{\perp}_\b(\o_n)= {1\over \o_n}\lim_{p\to 0}[-\la \hat
j_\pp^{P,\perp}; \hat j^{P,\perp}_{-\pp}\ra+\la
j^{D,\perp}_x\ra]\label{ccc}\ee
where, if $A=O_{\xx_1}\cdots O_{\xx_n}$, $\la A\ra= {\Tr[e^{-\b
H}{\bf T}(A)]\over \Tr[e^{-\b H}]}\Big|_T$, with ${\bf T}$ being
the time order product and $T$ denotes truncation. An analogous
definition holds for the parallel conductivity.

In order to compute the conductivity \pref{ccc} it is convenient
to introduce a Grassmann integral representation for the
correlation; we introduce the following {\it generating
functional}
\be e^{\WW_{t,\l}(A^{\perp})}=\int
P(d\psi)e^{-\VV(\psi)-B(A^{\perp},\psi)}\label{ccx1} \ee
where $\psi^\pm_{\xx,i}$ are Grassmann variables, $\xx=(x_0,x)$,
$P(d\psi)$ is the Grassmann integration with propagator $\d_{i,j}
\hat g(\kk)$
\be \hat g(\kk)={1\over -ik_0+\cos k-\cos p_F} \ee
with $k_0={2\pi\over\b}(n+{1\over 2})$, $\VV$ is the interaction
given by, if $\int d\xx=\int_{-\b/2}^{\b/2} dx_0\sum_x$
\bea &&\VV(\psi)=-\l\sum_{i=1}^2\int d\xx d\yy v(\xx-\yy)
\psi^+_{\xx,i}\psi_{\xx,i}\psi^+_{\yy,i}\psi_{\yy,i}\\
&&-\n\sum_{i=1}^2\int d\xx \psi^+_{\xx,i}\psi_{\xx,i} -t\int d\xx
(\psi^+_{\xx,1}\psi^-_{\xx,2}+\psi^+_{\xx,2}\psi^-_{\xx,1})\nn\eea
with $v(\xx-\yy)=\d(x_0-y_0) v(x-y)$ and $\n$ is a counterterm
which is introduced to take into account the renormalization of
the chemical potential; moreover \be B(\psi,A^{\perp})=-i\int d\xx
A^{\perp}_\xx[
\psi^+_{\xx,1}\psi^-_{\xx,2}-\psi^+_{\xx,2}\psi^-_{\xx,1}]\nn \ee
%
Defining $H_t(\xx)$ the Fourier transform of $\la\hat j^\perp_\pp
;\hat j^\perp_\pp\ra$ we can write
\be H_t(\xx)=t^2{\partial^2\over \partial A^{\perp}_\xx\partial
A^{\perp}_{\bf 0}}\WW_{t,\l}(A^{\perp})|_{0} \ee
%

The analysis of the functional integral \pref{ccx1} will be done
by Renormalization Group (RG), integrating smaller and smaller
momentum scales. The hopping $t$ introduces an intrinsic scale in
the RG analysis. For scales greater than the (renormalized)
hopping, it is natural to treat the hopping as a perturbation
using the chain variables $\psi^\pm$; on the other hand at smaller
scales the hopping cannot be considered a perturbation and it is
convenient to use the variables
\be \hat b_{\kk,1}={1\over \sqrt{2}}[\hat
\psi_{\kk,1}+\hat\psi_{\kk,2}]\quad\quad \hat b_{\kk,2}={1\over
\sqrt{2}}[\hat\psi_{\kk,1}-\hat\psi_{\kk,2}]\label{cc} \ee
in terms of which the free action is diagonal but the
$b^\pm_{\kk,1}, b^\pm_{\kk,2}$ have different Fermi momentum. Note
that the temperature acts an infrared cut-off so that for
temperatures not too small only the first regime is present.

The first step of the RG analysis is the decomposition of the
propagator $g(\kk)$ as a sum of propagators supported close to the
two Fermi points $\pm p_F$ and more and more singular in the
infrared region, labelled by a quasi particle index $\a=\pm$
(labelling the Fermi points) and by an integer $h\le 0$:

\be \hat g(\kk)=\hat g^{(1)}(\kk)+\sum_{h= h_\b}^0 \sum_{\a=\pm}
\hat g_{\a}^{(h)}(\kk-\pp_F^\a) \ee
with $\pp_F^\a=(0,p_F)$, $g_\a^{(h)}$ supported on $2^{h-1}\le
|\kk-\pp_F^\a|\le 2^{h+1}$ and $g^{(1)}(\kk)$ has support far from
the Fermi points. Note that $2^{h_\b}\sim \pi/\b$; the fact that
the temperature is finite implies that there is a finite number of
scales.

The RG integration procedure is defined recursively in the
following way. Assume that we have integrated the fields
$\psi^{(1)}_{i},\psi^{(0)}_{i,\a},...,\psi^{(h+1)}_{i,\a}$; we get
\bea &&e^{\WW_{t,\l}(A^{\perp})}=e^{F_h(A^{\perp})}\\
&&\int P(d\psi^{(\le h)})e^{-\VV^{(h)}(\sqrt{Z_h}\psi^{(\le
h)})-B^{(h)}(A^{\perp},\sqrt{Z_h}\psi^{(\le h)})}\label{ccx3}\nn
\eea
where $P(d\psi^{(\le h)})$ is the Grassmanian quadratic
integration with propagator given by
\be g^{(\le h)}_\a(\xx)={1\over\b  L}{1\over Z_{h}}\sum_\kk
e^{i\kk\xx} {\chi_h(\kk)\over -i k_0+\cos p_F-\cos (k+\a
p_F)}\label{mama}\ee
with $\chi_h(\kk)$ is a smooth cut-off function with support
$|\kk-\a {\bf p}_F|\le 2^{h+1}$ and $Z_h$ is the wave function
renormalization. The single scale propagator $g^{(h)}_\a(\xx)$ is
obtained from $\hat g^{(\le h)}_\a(\kk)$ replacing $\chi_h(\kk)$
with $f_h(\kk)$ with support $2^{h-1}\le |\kk'|\le 2^{h+1}$,
$\kk=(k_0,k')$, $k=k'+\a p_F$, $k'$ is the momentum measured from
the Fermi point. It can be written as
\be g^{(h)}_\a(\xx)=e^{i\a p_F x}{1\over\b  L}{1\over
Z_{h}}\sum_\kk e^{i\kk'\xx} {f_h(\kk')\over -i k_0+\a v_F
k'}+r^{(h)}(\xx) \label{mama1}\ee
with $r^{(h)}(\xx)$ with the same decay properties as
$g^{(h)}_\a(\xx)$ with an extra factor $2^h$; therefore, the more
we are close to the Fermi momenta ({\it i.e.} $-h$ is large), the
more $r^{(h)}(\xx)$ is a small correction and the propagator is
essentially coinciding with the one of a massless Dirac particle.
Finally $\VV^{(h)}$ is the {\it effective potential} expressed by
a sum of monomials of order $n$ in the fields $\psi^{(\le h)}$
multiplied times a kernel $W_{n,0}^{(h)}$, while $B^{(h)}$ is sum
of monomials of order $n$ in $\psi$ and $m$ in $A^{(\perp)}$ with
kernels $W_{n,m}^{(h)}$. According to power counting, using that
$\hat g_\a(\kk)^{(h)}\sim 2^{-h}$ and $\int d\kk \hat
g^{(h)}_\a(\kk)\sim 2^{h}$, the "naive" scaling dimension of such
monomials is
\be D=2-n/2-m\label{dim} \ee
In the RG analysis we have to decompose $\VV^{(h)}$ (a similar
decomposition must be done also for $B^{(h)}$) as
\be \VV^{(h)}=\LL \VV^{(h)}+\RR\VV^{(h)} \ee
with $\RR=1-\LL$; $\LL \VV^{(h)}$ is the {\it relevant} or {\it
marginal} part of the effective interaction while $\RR\VV^{(h)}$
is the {\it irrelevant} part. Generally this decomposition is
dictated by the naive scaling dimension \pref{dim}; $\LL$ should
select the terms with positive or vanishing dimension $D$.
However, if the temperature verifies the condition
\be 2^{h_\b}>t_{h_\b}\label{21} \ee
where $t_h$ is the hopping at scale $h$, there is an improvement
with respect to naive power counting, and certain terms which are
dimensionally relevant or marginal are indeed irrelevant. In order
to verify this fact, we can split the kernels as $
W_{n,m}^{(h)}=W_{n,m}^{(a)(h)}+W_{n,m}^{(b)(h)}$ where
$W_{n,m}^{(a)(h)}$ is obtained from $W_{n,m}^{(h)}$ setting $t=0$.
In the case $n=4,m=0$ (with vanishing scaling dimension)
\be \LL \hat W_{4,0}^{(h)}(\underline\kk')=\hat
W_{4,0}^{(a)(h)}(\underline{\bf 0})\label{gggg} \ee
so that
\be \RR \hat W_{4,0}^{(h)}(\underline\kk')=[\hat
W_{4,0}^{(a)(h)}(\kk')-\hat W_{4,0}^{(a)(h)}(\underline{\bf
0})]+\hat W_{4,0}^{(b)(h)}(\kk')\label{hh} \ee
The first term in the r.h.s. of \pref{hh} can be rewritten as
$\underline\kk'\cdot\underline{\bf \partial}W_{n,m}^{(a)(h)}$, and
this produces an improvement $\sim 2^{h'-h}$ in the bound of the
kernel, if $h'$ is the scale of the momentum, which is sufficient
to make it irrelevant. Similarly the second term in \pref{hh},
namely $\hat W_{4,0}^{(b)(h)}(\hat\kk')$, has an extra $t_h
2^{-h}\le 2^{h_\b-h}$ with respect to the bound for
$W_{4,0}^{(h)}$, which again is enough to make it irrelevant;
therefore, the true marginal contribution is given by the r.h.s.
of \pref{gggg}. Therefore the only marginal quartic terms involve
fermions with the same chain index, and that the corresponding
effective coupling coincide with the one of the uncoupled $t=0$
case.

Similarly we define, for the terms with $n=2$ and the same chain
index
\be \LL \hat W_{2,0}^{(h)}(\underline\kk')=\hat
W_{2,0}^{(a)(h)}(\underline{\bf 0})+\kk'\partial \hat
W_{2,0}^{(a)(h)}(\underline{\bf 0})\ee
Note that $W_{2,0}^{(b)(h)}$ has an extra $(t_h 2^{-h})^2$ (there
are no terms linear in $t_h$ by conservation of the chain index).
Finally, if $n=2,m=0$ and the fermionic fields have different
chain index
\be \LL \hat W_{2,0}^{(h)}(\underline\kk')=\hat
W_{2,0}^{(h)}(\underline{\bf 0})\ee
Note that the terms with $n=2$ and an extra derivative are
irrelevant as they have at least an extra ${t_h 2^{-h}}$.
Therefore
\bea&&\LL\VV^{(h)}(\psi)=\sum_{i=1}^2 [\l_h\int d\xx[
\psi^+_{\xx,i,+}\psi^-_{\xx,i,+}\psi^+_{\xx,i,-}\psi^-_{\xx,i,-}\nn\\
&&2^h\sum_{\a=\pm}
\n_h\psi^+_{\xx,i,\a}\psi^-_{\xx,i,\a}+\d_h\sum_{\a=\pm}
\psi^+_{\xx,i,\a}\psi^-_{\xx,i,\a}]+\\
&&+t_h\sum_{\a=\pm}\int d\xx
(\psi^+_{\xx,1,\a}\psi^-_{\xx,2,\a}+\psi^+_{\xx,2,\a}\psi^-_{\xx,1,\a})]\nn
\eea
In the above expression $\l_h$ represents the effective
interaction at momentum scale $h$, $\d_h$ the effective Fermi
velocity, $\n_h$ the shift of the chemical potential and $t_h$ the
effective hopping. By definition, $Z_h,\l_h,\n_h,\d_h$ are {\it
exactly the same} as in the theory with $t=0$. It is possible to
choose $\n$ so that $\n_h$ remain small for any $h$. By combining
Ward-Identities at each Renormalization group iteration together
with Schwinger-Dyson equation it follows, see \cite{BM}, that \be
\l_h\to_{h\to-\io} \l_{-\io}(\l)\quad \d_h\to_{h\to-\io}
\d_{-\io}(\l) \ee with $\l_{-\io}(\l),\d_{-\io}(\l)$ analytic
functions of $\l$; moreover  \be Z_h\sim 2^{-\h h} \ee with $\h$
analytic in $\l$ and $\h=a\l^2+O(\l^3)$ with $a>0$. Moreover, in
\cite{BM} (and references therein) it is also proven that kernels
$W^{(h)}_{n,m}$ are {\it analytic functions} of
$\{\l_k,\n_k,\d_k,t_k\}_{k\ge h}$: analyticity (implying the
"non-perturbative" nature of the method) is a very non trivial
property obtained exploiting anticommutativity properties of
Grassmann variables, via {\it Gram inequality} for determinants
and Bridges-Battle-Federbush formula for truncated expectations.

Regarding the flow of $t_h$ we obtain
\be t_{h-1}={Z_h\over Z_{h-1}}(t_h+\b_t^{(h)})\label{sol11} \ee
with $|\b_t^{(h)}|\le C_1 t_h \l^2(t_h 2^{-h})^2$. It is easy to
see by induction that $|Z_h t_h-t|\le C_2 t |\l|$. Assume indeed
that it is true for $k\ge h$; therefore for $\l,t$ small enough
\be |t_{h-1} Z_{h-1}-t|\le  2 t C_1\l^2\sum_{k=h}^0(t_k
2^{-k})^2\ee
from which, using \pref{21}, the inductive assumption follows.
Note that the effective hopping, even if {\it relevant} in the RG
sense according to naive power counting, remains small in this
region of temperatures. Moreover, from \pref{21} we obtain the
condition between the temperature and the hopping
\be \b^{-1}\ge t^{1\over 1-\h}(1+O(\l^2))\label{jjjj}\ee

Regarding the effective source $B^{(h)}$, we define \be \LL
W^{(h)}_{2,1}(\kk',\pp)=W^{(a)(h)}_{2,1}({\bf 0},{\bf
0})=1\label{bo}\ee Indeed the graphs contributing to
$W^{(a)(h)}_{2,1}({\bf 0})$ are one particle reducible (as the
interaction involves only fermions from the same chain) and
$g^{(h)}(\kk')|_{\kk'=0}=0$. Therefore (assuming that
$A^{\perp}_\pp$ has small support around $\pp=0$)
\bea &&\LL B^{(h)}(A^{\perp},\sqrt{Z_h}\psi)=\\
&&-i\sum_{i=1}^2\sum_{\a=\pm}\int d\xx
A_{\xx}^{\perp}(\psi^+_{\xx,1,\a}\psi^-_{\xx,2,\a}+\psi^+_{\xx,2,\a}\psi^-_{\xx,1,\a})\label{m1}
\nn\eea
As the flow of the effective parameters corresponding to the
relevant and marginal operators is bounded, the following bound is
obtained, for $\b$ verifying \pref{jjjj}
\be {1\over L\b}\int d\underline\xx |W^{(h)}_{n,m}(\xx)|\le C
2^{h(2-{n\over 2}-m)}\label{ggvv}
\ee
In order to compute the conductivity we have to separate the terms
proportional to $t^2$ in both the paramagnetic and diamagnetic
contributions to \pref{ccc} from the rest. We write the
current-current correlation as

\be t^{-2} H_t(\xx)={\partial^2 \WW_{0,\l}\over
\partial A^\perp_{\xx}\partial A^\perp_{\bf 0}}|_0+t^{-2}\tilde H_t(\xx)\label{bbb} \ee
where the first term in the r.h.s. is independent from $t$ and,
from \pref{ggvv}
\be \int d\xx |\tilde H_t(\xx)|\le C t^2\sum_{h=h_\b}^0 ({t\over
2^{h}})^2 Z_h^{-4}\le 2t^2 C(t \b^{1-2\h})^2 \ee
In order to compute the conductivity we still have to compute
$<j^{D,\perp}_x>$; introducing the generating functional
\be e^{\bar\WW_{t,\l}(J)}=\int P(d\psi)e^{-\VV(\psi)-t\int d\xx
J_\xx h_\xx}\label{ccx} \ee
where
$h_\xx=\psi^+_{\xx,1}\psi^-_{\xx,2}+\psi^+_{\xx,2}\psi^-_{\xx,1}$
we get
\be <j^{D,\perp}_x>={\partial \bar\WW_{t,\l}\over
\partial J_\xx}|_0=
t^{2}\int d\xx_1 {\partial^{2}\bar\WW_{0,\l}\over
\partial J_\xx\partial J_{\xx_1}}|_0+\D
\ee
where
\be \D=\sum_{n=3}^\io {t^{n+1}\over n!}\int d\xx_1...\int d\xx_n
{\partial^{n+1}\bar\WW_{0,\l}\over \partial J_\xx\partial
J_{\xx_1}...\partial J_{\xx_n}}|_0 \ee
and only $n$ odd contribute. From the analogue of \pref{ggvv} the
l.h.s. is bounded by the sum over $h$ of $\sum_{n=3}^\io t^{n+1}
2^{-h(n-1)}Z_h^{-4}$ so that, for $t\b$ small
\be |\D|\le C_1 t^2\sum_{h=h_\b}^0 (2^{- h}t Z_h^{-2})^2 \le C_2^n
t^2 (t\b^{1-2\h})^2\ee
Note finally that
\be -{\partial^2 \WW_{0,\l}\over
\partial A^\perp_{\xx}\partial A^\perp_{\bf 0}}|_0+
\int d\xx_1 {\partial^{2}\bar\WW_{0,\l}\over
\partial J_\xx\partial J_{\xx_1}}|_0\ee
can be rewritten as \bea&& \la
\psi^+_{\xx,1}\psi^-_{\xx,2}-\psi^+_{\xx,2}\psi^-_{\xx,2};
\psi^+_{\xx,1}\psi^-_{\xx,2}-\psi^+_{\xx,2}\psi^-_{\xx,2}\ra_{0,\l}+\nn\\
&&\int d\yy\la
\psi^+_{\xx,1}\psi^-_{\xx,2}+\psi^+_{\xx,2}\psi^-_{\xx,2};\psi^+_{\yy,1}\psi^-_{\yy,2}+\psi^+_{\xx,2}\psi^-_{\yy,2}\ra_{0,\l}\nn
\eea or equivalently

\bea&&
\la \psi^+_{\xx,1}\psi^-_{\xx,2}-\psi^+_{\xx,2}\psi^-_{\xx,2};\psi^+_{\xx,1}\psi^-_{\xx,2}-\psi^+_{\xx,2}\psi^-_{\xx,2}\ra_{0,\l}-\nn\\
&&\int
d\yy\la\psi^+_{\xx,1}\psi^-_{\xx,2}-\psi^+_{\xx,2}\psi^-_{\xx,2};\psi^+_{\yy,1}\psi^-_{\yy,2}-\psi^+_{\xx,2}\psi^-_{\yy,2}\ra_{0,\l}\nn
\eea
This means that there is an important cancellation between the
paramagnetic and diamagnetic part of the conductivity; indeed
\bea &&-\la j_\pp^{P,\perp}
;j^{P,\perp}_{-\pp}\ra+\la j^{D,\perp}_x\ra=\\
&&t^2 \int d\xx (e^{i\o_n x_0}-1)\la
j_{\xx,D}^{\perp}j_{\yy,D}^{\perp}\ra_{0,\l}+O(t^2
(t\b^{1-2\h})^2)\label{tt}\nn \eea
where
\be \la
j_{\xx,D}^{\perp};j_{\yy,D}^{\perp}\ra_{0,\l}=\la\psi^-_{\xx,1}
\psi^+_{\yy,1}\ra_{0,\l}\la\psi^+_{\yy,2}
\psi^-_{\xx,2}\ra_{0,\l}\nn \ee
Therefore for $\o_n$ small \bea && |\int d\xx (e^{i \o_n
x_0}-1)\la j_{\xx,D}^{\perp}j_{\yy,D}^{\perp}\ra_{0,\l}|\nn\\
&& \le C_1\int_{|\xx|\le |\o_n|^{-1}} d\xx {|x_0 \o_n|\over
1+|\xx|^{2+2\h}}+\\
&&C_1\int_{|\xx|\ge |\o_n|^{-1}} d\xx {1\over 1+|\xx|^{2+2\h}}\le
{C_2\over \h} |\o_n|^{2\h}\nn \eea

In conclusion the transverse conductivity is given by \pref{x1},
for $t<<\b^{-1}<< \o_n<<1$. In the non interacting case
$t^{-2}\o_n\s_{\b}^\perp(\o_n)\sim {2\over\pi \sin p_F}$ so that
we can conclude that the presence of the inter-chain interaction
decreases the transverse conductivity in this regime.

The transverse conductivity should be compared with the parallel
conductivity $\s_{\b}^\parallel$, defined as in \pref{ccc} with
$j^{D,\parallel},j^{P,\parallel}$ replacing
$j^{D,\perp},j^{P,\perp}$. For $t<<\b^{-1}<< \o_n<<1$ one gets
\be \o_n\s_{\b}^\parallel(\o_n)\sim 2 {v_F K\over\pi}\label{46a}
\ee
where $v_F=\sin p_F+O(\l)$ is the interacting Fermi velocity and
$K$ is the Luttinger liquid parameter $2\h=K+K^{-1}-2$. The above
formula, which is part of the Haldane's Luttinger liquid
conjecture \cite{H}, can be derived by a Renormalization Group
analysis analogue to the one described above, see \cite{BM1}. The
parallel current-current correlation is obtained by a generating
functional similar to \pref{ccx1} in which $B(A^{\perp},\psi)$ is
replaced by $B(A^{\parallel},\psi)=\int d\xx A^{\parallel}_\xx
j_\xx^{P,\parallel}$; after the integration of the fields
$\psi^{(1)}_{i},\psi^{(0)}_{i,\a},...,\psi^{(h+1)}_{i,\a}$ we get
an expression similar to \pref{ccx3} with
$B^{(h)}(A^{\perp},\sqrt{Z_h}\psi)$ replaced by
$B^{(h)}(A^{\parallel},\sqrt{Z_h}\psi)$ with
\be \LL B^{(h)}(A^{\parallel},\sqrt{Z_h}\psi)=Z^{(1)}_h \int d\xx
j^{P,\parallel(\le h)}_\xx\ee
and
\be {Z^{(1)}\over Z_h}=1+O(\l)\label{jj}\ee as a consequence of a
Ward identity.  Therefore the renormalization of the parallel
current is proportional to the wave function renormalization
(while there is no renormalization of the transverse current, see
\pref{bo}) and this explains why anomalous power law exponents do
not appear in the frequency dependence of the parallel
conductivity. In conclusion, from \pref{x1} and \pref{46a} we see
that $\h$ can be independently determined in experiments on two
chain systems either from the amplitude of the parallel
conductivity or from the exponent in the orthogonal conductivity.

As a final remark, we stress that the above analysis is true only
for temperatures greater than $\sim t^{1\over 1-\h}$; at lowest
temperature the RG analysis would be identical to the previous one
up to scale $t^{1\over 1-\h}$, but at lowest scales one should
perform the change of variables \pref{cc}; the system would be
described in terms of two fermions with different Fermi momenta
(the difference is $O(t^{1\over 1-\h})$. In this second regime the
power counting improvement described above is not valid, and this
produces several (not a single one, as in spinless Luttinger
liquids) effective quartic couplings with a generically unbounded
RG flow.


\section{Bilayer graphene}

An analysis similar to the previous one can be repeated for a
model of bilayer graphene, described in terms of electrons on the
honeycomb lattice interacting through an $U(1)$ quantized gauge
field, which can represent either the e.m. interaction or the
effects of ripples or disorder, see {\it e.g.} \cite{V2}.

We introduce creation and annihilation fermionic operators
$\psi_{\vec x,i}^\pm=(a^\pm_{\vec x,i}, b^\pm_{\vec x + \vec
\d_1,i})= |\BBB|^{-1}\int_{\vec k\in\BBB}d\vec k\,\psi^\pm_{\vec
k,i,\s} e^{\pm i\vec k\vec x}$ for electrons with plane index
$i=1,2$ and sitting at the sites of the two triangular sublattices
$\L_A$ and $\L_B$ of a honeycomb lattice; we assume that $\L_A$
has basis vectors $\vec l_{1,2}= \frac12(3,\pm\sqrt3)$ and that
$\L_B=\L_A+\vec\d_j$, with $\vec \d_1=(1,0)$ and
$\vec\d_{2,3}=\frac12 (-1,\pm\sqrt3)$ the nearest neighbor
vectors; $\BBB$ is the first Brillouin zone and
$|\BBB|={8\pi^2\over 3\sqrt{3}}$. In the absence of e.m.
interaction, the Hamiltonian is
\be H=H_1+H_2+P \ee
where
\be
H_i=-\sum_{\substack{\vec x\in\L_A \\
j=1,2,3}}a^{+}_{\vec x,i} b^{-}_{\vec x + \vec \d_j,i}+ c. c. \ee
describes the hopping of fermions in the plane and
\bea &&P=-t\sum_{\vec x\in\L_A} [a^{+}_{\vec x,1}
a^{-}_{\vec x ,2}+a^{+}_{\vec x,2}  a^{-}_{\vec x ,1}+\nn\\
&& b^{+}_{\vec x+\vec \d_1,1} b^{-}_{\vec x+\vec \d_1
,2}+b^{+}_{\vec x+\vec \d_1,2}  b^{-}_{\vec x+\vec \d_1 ,1}]\eea
describes the fermionic hopping from one plane to another; either
$e$ and $t$ will be assumed small. The interaction with a
transverse classical e.m. field is introduced via the Peierls
substitution. If $A^\perp$ is a classical external field
\bea &&P=-t\sum_{\vec x\in\L_A} [a^{+}_{\vec x,1} e^{ i
A^{\perp}_{\vec x}}a^{-}_{\vec x ,2}+a^{+}_{\vec x,2} e^{-i
A^{\perp}_{\vec
x}} a^{-}_{\vec x ,1}+\nn\\
&& b^{+}_{\vec x+\vec \d_1,1} e^{i A^{\perp}_{\vec x+\vec\d_1}}
b^{-}_{\vec x+\vec \d_1 ,2}+b^{+}_{\vec x+\vec \d_1,2} e^{-i
A^{\perp}_{\vec x+\vec\d_1}} b^{-}_{\vec x+\vec \d_1 ,1}]\eea
the paramagnetic and diamagnetic part of the transverse current
are
\bea &&j^{P,\perp}_{\vec x}={\partial H(A)\over\partial
A^{\perp}_x}|_0={t\over i}[a^{+}_{\vec x,1} a^{-}_{\vec x
,2}+\nn\\
&&b^{+}_{\vec x+\vec \d_1,1}b^{-}_{\vec x+\vec \d_1,2}-a^{+}_{\vec
x,2} a^{-}_{\vec x ,1}-b^{+}_{\vec x+\vec \d_1,2}b^{-}_{\vec
x+\vec \d_1,1}
\nn\\
&&j^{D,\perp}_{\vec x}={\partial^2 H(A)\over\partial^2
A^{\parallel}_x}=t[a^{+}_{\vec x,1} a^{-}_{\vec x
,2}+\nn\\
&&b^{+}_{\vec x+\vec \d_1,1}b^{-}_{\vec x+\vec \d_1,2}+a^{+}_{\vec
x,2} a^{-}_{\vec x ,1}+b^{+}_{\vec x+\vec
\d_1,2}b^{-}_{\vec x+\vec \d_1,1} \nn\\
\eea
and the transverse conductivity is defined as in \pref{ccc}
divided by ${3\sqrt{3}\over 2}$, the area of the hexagonal cell of
the honeycomb lattice.

We assume now that the electrons interact through an $U(1)$ gauge
field; the current-current correlation is obtained from the
following generating functional
\be e^{\WW_{t,e}(A^{\perp})}=\int P(d\psi) P(d
A)e^{-\VV(\psi,A)-B(A^{\perp},\psi)}\label{ccx} \ee
where $\psi=(a,b)$ a couple of Grassmann variables (with slight
abuse of notation, the Grassmann and the fermionic operators are
denoted with the same symbol), $P(d\psi)$ is the fermionic
gaussian integration for $\psi^\pm_{\kk,i}$ ($i=1,2$ denotes the
plane), $\kk=k_0,\vec k$, $k_0={2\pi\over\b}(n+{1\over 2})$, with
propagator $\d_{i,j} g(\kk)$ with
\be g^{-1}(\kk)= -\left(\begin{array}{cc} ik_0 & v_0 \O^*(\vec k)
\\ v_0 \O(\vec k) & ik_0 \end{array}\right)\;,\label{1.8}\ee
and $v_0=\frac32 $ and $\O(\vec k)=\frac23\sum_{j=1}^3 e^{i\vec
k(\vec\d_j-\vec\d_1)}$. The complex dispersion relation $\O(\vec
k)$ vanishes only at the two {\it Fermi points} $\vec p_{F}^{\
\pm}=(\frac{2\pi}{3},\pm\frac{2\pi}{3\sqrt{3}})$ and close to them
assumes the form of a relativistic dispersion relation $\O(\vec
p_F^\pm+\vec k')\simeq i k_1'\pm k_2'$. Moreover
\bea &&\VV=-\int d\xx [a^{+}_{\xx,i,\s} b_{\xx + {\bf \d}_j,\s}
e^{ie\int_0^1\vec\d_j\cdot\vec A_i(\xx+s{\bf\d}_j)\,ds}
 + c. c.]+\nn\\
&&\int d\xx A^{(0)}_{\xx,i}
a^+_{\xx,i}a_{\xx,i,\s}+A^{(0)}_{\xx+{\bf\d}_1,i}
b^+_{\xx+{\bf\d}_1,i}b_{\xx+{\bf\d}_1,i} \eea
where $\int d\xx\equiv\sum_{x\in \L_A}\int dx_0$ and
$A_{\m,i}=(\vec A_i,A^{(0)}_i)$ is a boson field with propagator
$\d_{i,j}w(\pp)$ with
\be w(\pp)={\chi(\sqrt{\o_n^2+c^2 p^2})\over \sqrt{\o_n^2+c^2 p^2}
}\label{kkl}
 \ee
where $\chi$ is a cut-off function forbidding momenta either too
large and smaller than the temperature. Finally the source term is
given by
\be B(A^{\perp},\psi)=\int d\xx A^{\perp}_\xx j^{P,\perp}_\xx\ee
We proceed exactly as in the previous case, writing the photon
propagator as sum of propagator more and more singular in the
infrared region, and the fermionic propagator as a sum of
propagators supported close to the two Fermi points $\vec
p_F^\pm=(\frac{2\pi}{3},\pm\frac{2\pi}{3\sqrt{3}})$, labelled by a
quasi particle index $\a=\pm$ (labelling the Fermi points) and by
an integer $h\le 0$:
\bea &&w(\pp)=\sum_{h=h_\b}^1 w^{(h)}(\pp)\\
&& g(\kk)=g^{(1)}(\kk)+\sum_{h= h_\b}^0 \sum_{\a=\pm}
g_{\a}^{(h)}(\kk-\pp_F^\a)\nn \eea
with $w^{(h)}(\pp)$ supported in $2^{h-1}\le |\pp|\le 2^{h+1}$,
$g_\a^{(h)}$ supported on $2^{h-1}\le |\kk-\pp_F^\a|\le 2^{h+1}$
and $g^{(1)}(\kk)$ has support far from the Fermi points.

Assume that we have integrated out the fields
$\psi^{(1)},..,A^{(h+1)},\psi^{(h+1)}$, $h\ge h_ \b$ so that
\bea &&e^{W(A^{\perp})}= e^{F_h(A^\perp)}\\
&&\int P(d\psi^{(\le h})\int P(dA^{(\le
h)})e^{\VV^{(h)}(A,\sqrt{Z_h}\psi)+B_h(A^\perp,\sqrt{Z_h}\psi)}\nn
\eea
where $P(dA^{(\le h)})$ is the gauge field integration with
propagator $\d_{i,j}\d_{\m,\n}w^{\le h}(\pp)$, with $w^{\le
h}(\pp)=\sum_{k=-\io}^h w^{(k)}(\pp)$, while $P(d\psi^{(\le h})$
is the integration of the fermionic field $\psi_{i,\a}$ with
propagator $\d_{i,i'}\d_{\a,\a'} g^{\le h}_\a(\kk-\pp_F^{(\a)})$
with, if $\kk'=\kk-\pp_F^{(\a)}$, $g_{\a}^{(\leq h)}(\kk')=$
\be =\frac{\chi_h(\kk')}{Z_h}\left(\begin{array}{cc} -ik_0 & v_h(i
k_1'-\a k_2')\\ v_h(-ik_1'-\a k_2') & -ik_0 \end{array}
\right)^{\!\!\!-1}\!\!\!(1+R^{(h)}_{\o})\;.\label{3}\ee
In Eq.(\ref{3}) $\chi_h(\kk')$ is a cut-off function with support
in $|\kk'|\le 2^h$ and $|R^{(h)}_{\o}(\kk')|\le C |\kk'|^\th$ for
some $\th>0$, while $Z_h$ and $v_h$ are, respectively, the
effective wave function renormalization and Fermi velocity on
scale $h$.

The effective potential $\VV^{(h)}+B^{(h)}$ expressed by a sum of
monomials of order $n$ in the fields $\psi^{(\le h)}$, $m$ in
$A^{(\m)(\le h)}$ and $l$ in $A^{\perp}$, multiplied by kernels
$W^{(h)}_{n,m,l}$. According to power counting the naive scaling
dimension of such monomials is
\be D=3-n-m-l \label{dim1} \ee
Again there is a dimensional improvement with respect to power
counting if we are in a range of temperatures larger than the
hopping, that is $2^{h_\b}>t_{h_\b}$ where $t_h$ is the hopping at
scale $h$. We can split the kernels as
$W_{n,m,l}^{(h)}=W_{n,m,l}^{(a)(h)}+W_{n,m,l}^{(b)(h)}$ where
$W_{n,m,l}^{(b)(h)}$ is obtained from $W_{n,m}^{(h)}$ setting
$t=0$. We define the $\LL$ operator in the following way
\be \LL \hat W^{(h)}_{2,1,0}(\kk')=\hat W^{(a)(h)}_{2,1,0}(0) \ee
Note indeed that the extra $t_h 2^{-h}\le 2^{h_\b-h}$ in
$W^{(b)(h)}_{2,1,0}(0)$ is sufficient to make it irrelevant.
Regarding the terms quadratic in the gauge fields, $\LL \hat
W^{(h)}_{0,2,0}(\pp)=\hat W^{(a)(h)}_{0,2,0}(0)+\pp\partial \hat
W^{(a)(h)}_{0,2,0}(0)$, where we have used that $\hat
W^{(b)(h)}_{0,2,0}(0)$ has an extra $(2^{-h}t_h)^2$ with respect
to the naive dimension; moreover either $\hat
W^{(a)(h)}_{0,2,0}(0)$ and $\partial\hat W^{(a)(h)}_{0,2,0}(0)$
are vanishing as consequence of the gauge symmetry, see
\cite{GMP2}. Finally the terms quadratic in the fermionic
variables, if they have the same plane index then $\LL \hat
W^{(h)}_{2,0,0}(\kk')=\hat W^{(a)(h)}_{2,0,0}(0)+\kk'\partial \hat
W^{(a)(h)}_{2,0,0}(\kk')$
where we have used that in $\hat W^{(b)(h)}_{2,0,0}$ there is an
extra gain $O((t_h 2^{-h})^2$, due to the conservation of the
plane index $i$. On the other hand for the quadratic terms with
different plane index \be \LL \hat W^{(h)}_{2,0,0}(\kk')=\hat
W^{(h)}_{2,0,0}({\bf 0}) \ee
Therefore
\bea &&\LL\VV^{(h)}(A,\psi)=t_h \int d\xx j_{\xx}^{D,\perp}+\\
&&\sum_{\m,i,\a}\bar{e}_{\m,h} \int
\frac{d\kk}{(2\pi)|\BBB|}\,\frac{d\pp}{(2\pi)^3}\,
\psi_{\kk+\pp,i,\a}^{+}
 \G_{\m}^\o\psi^{-}_{\kk,i,\a} A^{\m}_{i}(\pp)
 \nn\eea
where: $\bar e_{0,h}=e_{0,h}$, $\bar e_{i,h}=v_h e_{i,h}$,
$e_{1,h}=e_{2,h}$ (thanks to discrete rotational symmetry),
$\G_\m^\o:=\G_\m(\vec p_F^\o,\vec 0)$ (with $\G_0^\o=-iI$,
$\G_1^\o=-\s_2$, $\G_2^\o=-\o\s_1$ and $\s_{1,2}$ the first two
Pauli matrices) and $\RR\VV^{(h)}$ a sum of terms that are
irrelevant in the RG sense. By construction the flow of the
effective parameters is the same as in the model with $t=0$; it
was show in \cite{GMP2}, by a rigorous implementation of Ward
Identities in the RG scheme, that the effective charges flows to a
line of fixed points and the Fermi velocity increases up to the
light velocity
\be e_h\to e_{-\io}\quad\quad v_h\to c \ee
Moreover, the wave function renormalization $Z_h$ diverges with
anomalous exponents
\be Z_h\sim 2^{-\h h}\quad\quad \h={e^2\over 12\pi^2}+... \ee
Finally regarding the flow of $t_h$ we obtain
\be t_{h-1}={Z_h\over Z_{h-1}}(t_h+\b_t^{(h)})\label{sol} \ee
with $|\b_t^{(h)}|\le C_1 e^6 t_h [{t_h\over 2^h}]^2$, and again
by induction $|Z_h t_h-t|\le C_2 t e^6$. We assume that the
temperature verifies \pref{21} which implies $\b^{-1}\ge t^{1\over
1-\h}(1+O(e^2))$. Regarding the effective source $B^{(h)}$, we
define $\LL W^{(h)}_{2,0,1}(\kk',\pp)=W^{(a)(h)}_{2,0,1}({\bf
0},{\bf 0})=1$ as again the graphs contributing to
$W^{(a)(h)}_{2,0,1}({\bf 0})$ are one particle reducible and
$g^{(k)}(\kk')|_{\kk'=0}=0$. As the flow of the effective
parameters corresponding to the relevant and marginal operators is
bounded, the following bound is obtained, for $h\ge h_\b$ (order
by order in the renormalized expansion)
\be {1\over \L\b}\int d\underline\xx
|W^{(h)}_{n,m,l}(\underline\xx)|\le C
2^{h(3-n-m-l)}
\label{ggxx}\ee
Using the same notation as in \pref{bbb}
%
%
\be \int d\xx |x_0| |\tilde H_t(\xx)|\le C t^2\sum_{h=h_\b}^0
({t\over 2^{h}})^2 Z_h^{-4}\le 2t^2 C(t \b^{1-2\h})^2 \ee
Moreover, as in the previous case we introduce a generating
functional $\bar\WW_{t,e}(J)$ with source $t\int d\xx J_\xx h_\xx$
where $j^{D,\perp}_\xx=t h_\xx$
we get
\be \la j^{D,\perp}_x\ra = t^{2}\int d\xx_1
{\partial^{2}\bar\WW_{0,e}\over
\partial J_\xx\partial J_{\xx_1}}|_0+\D
\ee
From the analogue of \pref{ggxx} the l.h.s. is bounded by the sum
over $h$ of $\sum_{n=3}^\io t^{n+1} 2^{-h(n-2)}Z_h^{-4}$ so that,
for $t\b$ small
\be \o_n^{-1}|\D|\le t^2\sum_{h=h_\b}^0 \b 2^h [{ t2^{-h}\over
Z_{h}^2}]^{2}\le C t^2(t\b^{1-2\h})^2\ee
Note finally that
\be \la j^{P,\perp}_\xx; j^{P,\perp}_\yy\ra_{0,e}=\la
j^{D,\perp}_\xx;  j^{D,\perp}_\yy\ra_{0,e} \ee
and
\bea && |\int d\xx x_0 (e^{i \o_n
x_0}-1)\la j_{\xx,D}^{\perp}; j_{\yy,D}^{\perp}\ra_{0,\l}|\nn\\
&& \le C_1\int_{|\xx|\le \o_n^{-1}} d\xx |x_0| {|x_0 \o_n|\over
1+|\xx|^{4+2\h}}+\nn\\
&&C_1\int_{|\xx|\ge \o_n^{-1}} d^3\xx |x_0|{1\over
1+|\xx|^{4+2\h}}\le {C_2\over \h} |\o_n|^{2\h}\nn \eea
Therefore the conductivity in the interacting case is given by
\pref{x2} for $t<<\b^{-1}<< \o_n<<1$, that is
%
%
the transverse conductivity decreases with the frequency with the
anomalous exponent $2\h$. In absence of planar interaction $
t^{-2}\s^{\perp}_\b(\o_n)\sim {1\over 2}$, so that we can conclude
that the presence of planar long range interaction producing
Luttinger liquid behavior decreases the transverse conductivity.
Note also that the parallel conductivity does not display any
anomalous power law, as a consequence of a Ward Identity implying
the analogue of \pref{jj}, see \cite{GMP2}.

\section{Appendix: The non interacting case}

In the case of the two chain model if $\l=0$

\bea &&t^{-2}\o_n\s_\b^\perp(\o_n)=[\int d\kk g(\kk+\pp)
g(\kk)-\nn\\
&&\int d\kk g(\kk) g(\kk)]|_{p=0}+O((\b t)^2)\nn\eea
Note that
\be  \lim_{\b,L\to\io}{1\over L\b}\sum_\kk g(\kk+\pp)
g(\kk)|_{p=0}=0\label{aa} \ee
while
\be \lim_{\b,L\to\io}{1\over L\b}\sum_\kk g(\kk) g(\kk)={
2\over\pi \sin p_F} \ee
%

In the case of bilayer graphene, we get $t^{-2}\s_\b^\perp(\o_n)=$
\be {1\over\o_n}\int {d k_0\over (2\pi)}d\vec k
[F(\kk,\kk+\pp)-F(\kk,\kk)]|_{\vec p=0}+O((t\b)^2)\nn\ee
where
\bea &&
F(\kk_1,\kk_2)=2[g_{11}(\kk_1)g_{11}(\kk_2)+g_{22}(\kk_1)g_{22}(\kk_2)\nn\\
&&
+g_{12}(\kk_1)g_{12}(\kk_2)+g_{21}(\kk_1)g_{21}(\kk_2)]\label{zza}
\eea
The first term in the r.h.s. can be written as its value for
$\b=\io$ plus a rest $O(\b^{-1})$; the integral in the limit
$\b=\io$ can be decomposed in a part integrated in the region
$|\O(\vec k)|\le \e$ and $|\O(\vec k)|\ge \e$; the second term is
vanishing for $\o_n=0$ while in the first the contribution from
the first two terms in \pref{zza} are vanishing by parity, while
the rest gives ${1\over 2}$ at vanishing external frequency.

\vskip.3cm {\it Acknowledgements} The Author gratefully
acknowledges financial support from the ERC Starting Grant
CoMBoS-239694

\end{document}